\documentclass[10pt, conference, letterpaper]{IEEEtran}
\IEEEoverridecommandlockouts

\usepackage{cite}
\usepackage{amsmath,amssymb,amsfonts}
\usepackage[ruled,linesnumbered]{algorithm2e}
\usepackage{graphicx}
\usepackage{textcomp}
\usepackage{xcolor}
\usepackage{multirow}
\usepackage{makecell}
\usepackage{booktabs}
\usepackage{subcaption}
\usepackage[table]{xcolor}
\usepackage{colortbl}
\usepackage{pifont}
\usepackage{mathrsfs}
\usepackage{hyperref}
\hypersetup{hidelinks}

\newcommand{\cmark}{{\color{black}\ding{51}}}
\newcommand{\xmark}{{\color{black}\ding{55}}}

\def\BibTeX{{\rm B\kern-.05em{\sc i\kern-.025em b}\kern-.08em
    T\kern-.1667em\lower.7ex\hbox{E}\kern-.125emX}}
\definecolor{ours}{RGB}{211,211,211}

\begin{document}

\title{Pruned Traffic Trees: Native Semantic Compression with a Protocol-Structured Model Family for Encrypted Traffic Classification}

\author{
\IEEEauthorblockN{
Yuantu Luo\IEEEauthorrefmark{1},
Jun Tao\IEEEauthorrefmark{1},
Xiangyu Xu\IEEEauthorrefmark{2},
Linxiao Yu\IEEEauthorrefmark{1},
Kangying Li\IEEEauthorrefmark{1}
}
\IEEEauthorblockA{
\IEEEauthorrefmark{1}
\textit{School of Cyber Science and Engineering, Southeast University},
Nanjing, China\\
\{ytluo, juntao, yulinxiaoybbb, kangying\}@seu.edu.cn
}
\IEEEauthorblockA{
\IEEEauthorrefmark{2}
\textit{School of Computer Science and Engineering, Southeast University},
Nanjing, China\\
xy-xu@seu.edu.cn
}
}


\maketitle

\begin{abstract}

Deep learning has achieved strong performance in encrypted traffic classification (ETC), yet its computational cost limits deployment on resource-constrained network devices such as routers and middleboxes. 
Existing compression methods mainly operate on weights, channels, hidden representations, or predictions, but do not explicitly determine which protocol fields and structural contexts should remain. 
We propose Pruned Traffic Trees (PTT), a three-level protocol-structured model family that treats native protocol structures as compression units.
PTT-Full learns protocol-structured representations and field salience from complete Protocol Tree Graphs (PTGs), with flow-level self-supervised learning and protocol-presence-aware sparse execution. 
The learned salience and TopK+$k$ closure construct Distilled PTGs (PTG-Ds) for PTT-Distilled, while PTT-Lite inherits this topology and reduces width through structure-aligned transfer and flow-level logits distillation. 
Under flow-disjoint and Strong Information Information (SII)-masked settings, PTT-Full achieves Macro-F1 scores of 0.9519 and 0.9416 on CSTNET-TLS1.3 and CipherSpectrum, while PTT-Lite retains 0.9325 and 0.9136 with 80.3\% and 61.3\% fewer parameters, 98.85\% and 98.78\% lower effective GFLOPs, and 8.75$\times$ and 8.46$\times$ CPU inference speedups. 
These results demonstrate that treating protocol structure itself as the compression object enables effective performance-efficiency trade-offs for lightweight ETC. 
\end{abstract}

\begin{IEEEkeywords}
Encrypted Traffic Classification, Protocol-Structured Representation, Semantic Compression, Sparse Expert Model
\end{IEEEkeywords}

\section{Introduction}

Encrypted payloads obscure application content, forcing encrypted traffic classification (ETC) models to rely on the structural evidence that remains observable at network devices~\cite{intro_ETC1_survey1}. 
Packet dissection exposes protocol fields, their boundaries, and their positions in the protocol hierarchy, which are interpreted jointly rather than as isolated bytes during traffic analysis. 
Together, these properties make protocol fields natural semantic units for ETC: they support structure-aware representation and, unlike anonymous bytes or latent channels, can be explicitly retained or removed during compression. 
A field’s meaning depends on both its value and structural position, making field identity and protocol context inseparable. 
This suggests that ETC models should preserve protocol fields as persistent units throughout both representation learning and compression. 

Existing ETC methods commonly encode traffic using packet-direction and timing sequences ~\cite{Method_Sequence2_BAPM,Method_Sequence3_VarCNN,Method_Sequence4_DF,Method_Sequence5_TikTok}, raw-byte or token sequences ~\cite{Method_sequence9_deeppacket,Method_Sequence1_TMWF}, or image-like representations ~\cite{Method_Imagelike2_YaTC,Method_Imagelike1_RF}. 
These representations can achieve strong classification performance, but they do not explicitly preserve protocol-defined field boundaries and parent--child relations. 
Fixed-length processing may further introduce padding or truncation, which inserts artificial values or removes useful traffic information. Generic graph-based methods improve structural modeling, but their edges are often constructed from statistical or heuristic relations rather than the native hierarchy produced by packet dissection~\cite{Method_DGNN}. 
Consequently, protocol fields are rarely exposed as persistent, semantically identifiable units that can later be selected or removed during model compression.

Beyond representation accuracy, practical ETC deployment should satisfy the latency, memory, and computation budgets of routers and middleboxes. 
Existing lightweight designs typically compress models through pruning, quantization, width reduction, or knowledge distillation~\cite{Survey_GraphKD}. 
These methods operate mainly on weights, channels, hidden representations, or predictions, and therefore do not explicitly control which protocol evidence remains available to a compact classifier. 
Applying structure pruning directly to a graph does not fully resolve this issue either: independently retained nodes may become detached from the protocol paths that define their context. 
This suggests that lightweight ETC requires more than a smaller neural network. 
The representation itself should expose semantically meaningful units that can be removed while retaining the protocol context required by the remaining fields. 
In other words, compression should answer both \textbf{what protocol evidence to retain} and \textbf{how to retain it in a valid structural context}. 


To address these challenges, we present \textbf{Pruned Traffic Trees (PTT)}, a protocol-structured ETC model family built on Protocol Tree Graphs (PTGs)~\cite{PTGAMoE}. 
PTGs align parsed protocol fields and their parent--child relations with persistent graph nodes and edges. 
While the inherited PTG representation uses this structure for representation learning, PTT further turns its field-level alignment into an explicit compression interface. 
PTT-Full learns protocol-structured representations together with field salience from complete PTGs. 
The learned salience determines which fields are retained, while TopK+$k$ closure restores their required protocol paths to construct compact Distilled PTGs (PTG-Ds) and instantiate PTT-Distilled. 
PTT-Lite then inherits the compressed topology and reduces width through structure-aligned transfer and flow-level logits distillation. 
To ensure that the measured gains reflect transferable traffic patterns rather than shortcut indicators or flow leakage, all methods are evaluated under SII-masked~\cite{antileakage_sok_enigma} and flow-disjoint settings~\cite{antileakage_sok_sweet}. 

Different from model-level compression that operates on anonymous parameters or latent channels, PTT makes protocol fields and their native hierarchy explicit objects of compression. 
Overall, our contributions are summarized as follows: 
\begin{itemize}
    \item We propose \textbf{PTG-native structural compression}, which turns field-aligned PTGs into an explicit compression interface.
    Learned field salience determines which protocol fields are retained, while TopK+$k$ closure constructs the minimal hierarchy-closed PTG containing those fields.
    This reduces graph structure without breaking the native protocol context of the retained evidence.

    \item We develop \textbf{PTT-Full} as the high-capacity source model.
    It combines flow-level self-supervised learning with protocol-presence-aware sparse dispatch, executing experts only for present protocol components and converting protocol absence into inference acceleration. 

    \item We derive \textbf{PTT-Lite} from PTT-Distilled through structure-aligned width transfer and flow-level logits distillation.
    Experiments on two TLS~1.3 datasets under flow-disjoint and SII-masked settings show that the PTT variants provide strong performance--resource trade-offs. 
\end{itemize}

\section{Related Work}

\paragraph{Encrypted Traffic Representation and Classification}
Encrypted traffic classifiers encode flows as images~\cite{Method_Imagelike1_RF,Method_Imagelike2_YaTC}, raw-byte or token sequences~\cite{Method_Sequence3_VarCNN,Method_Sequence4_DF,related_byte1_RBLJAN}, packet-length and timing sequences~\cite{Method_Sequence1_TMWF,Method_Sequence2_BAPM}, or multi-level features~\cite{Method_Sequence5_TikTok,Method_Sequence6_AWF}.
Recent sequence models improve byte-level learning through joint byte--label attention or large-scale traffic pretraining~\cite{Method_Sequence7_ETBERT}. 
Graph methods capture packet-, message-, or feature-level interactions~\cite{related_graph1_DigTraffic}, while PTGAMoE aligns graph nodes and edges with parsed fields and their native hierarchy~\cite{PTGAMoE}. 
However, these representations either flatten protocol evidence into implicit features or model generic interactions without exposing protocol fields as explicit structural units for compression. 
Moreover, shortcut features and improper splitting can substantially inflate ETC results~\cite{antileakage_sok_enigma,antileakage_sok_sweet}.

\paragraph{Knowledge Distillation and Lightweight Traffic Analysis}
Existing lightweight traffic analysis mainly follows general model compression paradigms, including knowledge distillation, pruning, and representation reduction. 
Graph distillation transfers logits~\cite{GraphKD_Logits1_Freekd,GraphKD_Logits2_Boost,GraphKD_Logits3_T2-gnn}, embeddings~\cite{GraphKD_Embed1_SAIL,GraphKD_Embed2_AKD,GraphKD_Embed3_RDD}, cross-layer relations~\cite{GraphKD_Structure1_alignahead}, or neighborhood information to compact students~\cite{GraphKD_Structure2_CKD,GraphKD_Structure3_ColdBrew}.
Other lightweight traffic studies explore programmable-switch deployment, sample-efficient learning, graph cross-distillation, and contextual masking distillation~\cite{LightweightETA1_SentinelX,LightweightETA2_Trainwithone,LightweightETA3_SEADGAT,LightweightETA4_CMD}. 
However, these methods mainly compress model parameters or representations rather than explicitly preserving the protocol context of retained fields. 
Moreover, they neither select the protocol evidence to retain nor
preserve its native hierarchy. 

\paragraph{Mixture of Experts for Traffic Analysis} 
Sparse Mixture-of-Experts (MoE) models improve capacity--computation trade-offs through conditional expert activation~\cite{related_MoE2_Google,related_MoE5_survey}.
Recent traffic models use experts for sparse foundation modeling, heterogeneous traffic specialization, and multi-view fusion~\cite{MoE_ETC1_TrafficMoE1,MoE_ETC2_TrafficMoE2,MoE-ETC-TrafficMoE3}. 
However, their routing decisions are typically defined over learned tokens, modalities, or predefined branches, rather than protocol components physically present in each packet. 

PTT differs by treating parsed fields and their hierarchy as explicit semantic units for both sparse execution and compression. 
It leverages protocol presence as an explicit routing signal and further uses learned field salience to guide hierarchy-preserving protocol compression. 

\section{Protocol-Structured Source Model PTT-Full}
\label{sec:model_design}

\begin{figure*}[t]
    \centering
    \includegraphics[width=\linewidth]{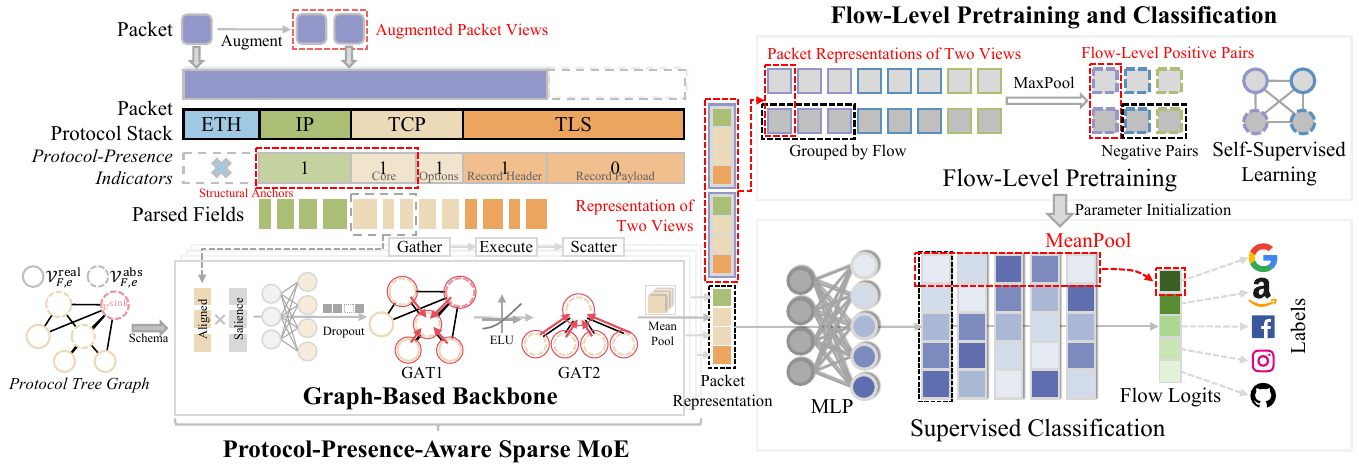}
    \caption{Overview of PTT-Full and its flow-level learning pipeline.}
    \label{fig:model_design}
\end{figure*}

As shown in Fig.~\ref{fig:model_design}, building on the Protocol Tree Graph (PTG) representation~\cite{PTGAMoE} and graph-expert backbone, PTT-Full adds flow-level self-supervised learning and protocol-presence-aware sparse dispatch. 

\subsection{Graph-Based Expert Backbone}
\label{subsec:backbone_formulation}

Let $\mathcal D=\{(F_i,y_i)\}_{i=1}^{N}$ be a labeled ETC dataset, where $F_i=(p_{i,1},\ldots,p_{i,T_i})$ is a flow and $y_i$ is its class.
For a maximum of $P$ packets per flow, $T_i=\min(N_i,P)$ retains the first packets in capture order.

Each packet is represented over an expert-indexed PTG set $\{\mathcal G_{F,e}\mid e\in\mathcal E\}$.
For expert $e$, the fixed Full schema is 
\begin{align}
    \mathcal G_{F,e}
    &= (\mathcal V_{F,e},\mathcal{E}_{F,e}),\\
    \mathcal V_{F,e}
    &= \mathcal V_{F,e}^{\mathrm{real}}
    \cup
    \mathcal V_{F,e}^{\mathrm{abs}},\\
    \mathcal{E}_{F,e}
    &= \mathcal{E}_{F,e}^{\mathrm{hier}}
    \cup
    \mathcal{E}_{F,e}^{\mathrm{sink}}.
\end{align}
Here, $\mathcal V_{F,e}^{\mathrm{real}}$ contains real protocol-field nodes whose values are obtained from packet dissection, whereas $\mathcal V_{F,e}^{\mathrm{abs}}$ contains abstract schema nodes used to preserve the protocol hierarchy.
$\mathcal{E}_{F,e}^{\mathrm{hier}}$ contains undirected hierarchical edges induced by the parent--child relations of packet dissection, while $\mathcal{E}_{F,e}^{\mathrm{sink}}$ connects schema nodes with the layer-wise sink node used for graph readout.
Each PTG is conceptually undirected. 
In implementation, every undirected edge is represented by two directed edges for bidirectional GAT message passing. 

For packet $p_{i,t}$, $\mathbf X_{i,t,e}\in\mathbb R^{|\mathcal V_{F,e}|\times d_F}$ contains node representations in the fixed schema order.
Each node $v$ has a shared gate logit $\alpha_{e,v}$ and activation
\begin{equation}
    s_{e,v}=\sigma(\alpha_{e,v}),
    \qquad 0<s_{e,v}<1.
    \label{eq:field_salience}
\end{equation}
The gated node representations and expert output are
\begin{align}
    \widetilde{\mathbf X}_{i,t,e}
    &=
    \left[
        s_{e,v}\operatorname{LN}(\mathbf x_{i,t,e,v})
    \right]_{v\in\mathcal V_{F,e}},
    \nonumber\\
    \mathbf h_{i,t,e}
    &=
    f_e(\widetilde{\mathbf X}_{i,t,e},\mathcal{E}_{F,e};\theta_{F,e}),
    \qquad
    \mathbf h_{i,t,e}\in\mathbb R^{d_F}.
    \label{eq:expert_mapping}
\end{align}
Here, $f_e$ is a two-layer graph attention encoder followed by mean readout:
\begin{align}
    \mathbf Z_{i,t,e}^{(0)}
    &=\operatorname{Dropout}(\widetilde{\mathbf X}_{i,t,e}),
    \nonumber\\
    \mathbf Z_{i,t,e}^{(1)}
    &=\operatorname{ELU}
    \left(
        \operatorname{GAT}_{e}^{(1)}
        (\mathbf Z_{i,t,e}^{(0)},\mathcal{E}_{F,e})
    \right),
    \nonumber\\
    \mathbf Z_{i,t,e}^{(2)}
    &=\operatorname{GAT}_{e}^{(2)}
    (\mathbf Z_{i,t,e}^{(1)},\mathcal{E}_{F,e}),
    \nonumber\\
    \mathbf h_{i,t,e}
    &=\operatorname{MeanPool}(\mathbf Z_{i,t,e}^{(2)}).
    \label{eq:expert_backbone}
\end{align}
The first layer uses multi-head attention and concatenation, while the second outputs dimension $d_F$. 
The learned real-node gates provide field salience for structural compression. 

Beyond the PTG schema and its corresponding graph-expert architecture, we introduce flow batching with variable lengths, flow-level self-supervised learning, and protocol-presence-aware sparse execution in PTT-Full to enhance the traffic representation. 

\subsection{Protocol-Presence-Aware Sparse MoE}
\label{subsec:sparse_moe}

\paragraph{Flow-Centric Batching}
To train and evaluate PTT at the flow granularity, packets from the
same flow are first organized into a variable-length micro-batch. 
The packet indices of flow $i$ form the micro-batch 
\begin{equation}
    \mathcal I_i^{\mathrm{pkt}}
    =\{(i,t)\mid 1\leq t\leq T_i\}.
    \label{eq:flow_packet_indices}
\end{equation}
For each optimization step, the sampler selects up to $N_{\mathcal B}$ flows with index set $\mathcal I_{\mathcal B}^{\mathrm{flow}}$ and concatenates their variable-length packet sets into a macro-batch: 
\begin{equation}
    \mathcal I_{\mathcal B}^{\mathrm{pkt}}
    =
    \biguplus_{i\in\mathcal I_{\mathcal B}^{\mathrm{flow}}}
    \mathcal I_i^{\mathrm{pkt}},
    \qquad
    |\mathcal I_{\mathcal B}^{\mathrm{pkt}}|
    \leq N_{\mathcal B}P.
    \label{eq:flow_centric_batch}
\end{equation}
This macro-batch introduces no packet padding, and flow identifiers are retained to regroup packet outputs for flow-level learning.

\paragraph{Protocol Presence}
A packet should not invoke an expert for a protocol component that it does not contain. 
For packet $p_{i,t}$ and expert $e$, define 
\begin{align}
    \mathcal P_{i,t,e}
    &=
    \{v\in\mathcal V_{F,e}^{\mathrm{real}}
    \mid v\text{ is observed in }p_{i,t}\},
    \nonumber\\
    m_{i,t,e}
    &=\mathbb I[\mathcal P_{i,t,e}\neq\varnothing].
    \label{eq:presence}
\end{align}
For the TCP-based flows considered here, IP and TCP-core experts are retained as structural anchors:
\begin{align}
    \mathcal E_{\mathrm{anchor}}
    &=\{e_{\mathrm{IP}},e_{\mathrm{TCPcore}}\},
    \nonumber\\
    m'_{i,t,e}
    &=\max(m_{i,t,e},\mathbb I[e\in\mathcal E_{\mathrm{anchor}}]).
    \label{eq:effective_presence}
\end{align}
where $\mathbb I[\cdot]$ denotes the indicator function. 
The active packet indices of expert $e$ are gathered into an expert-specific sub-batch
\begin{equation}
    \mathcal I_{\mathcal B,e}^{\mathrm{act}}
    =
    \{(i,t)\in\mathcal I_{\mathcal B}^{\mathrm{pkt}}
    \mid m'_{i,t,e}=1\}.
    \label{eq:expert_active_indices}
\end{equation}
Thus, protocol presence determines whether an expert can execute, while the fusion gate below determines how much its output contributes.

\paragraph{Sparse Dispatch and Fusion}
Masking an expert only after its forward pass does not reduce computation. 
PTT-Full therefore uses a gather--execute--scatter procedure: active packets are gathered into expert-specific sub-batches, each nonempty encoder executes once, and its outputs are scattered back to the original packet positions. 
In particular, 
\begin{equation}
    \mathcal I_{\mathcal B,e}^{\mathrm{act}}=\varnothing
    \quad\Longrightarrow\quad
    f_e\text{ is not executed}.
    \label{eq:expert_skip}
\end{equation}
Inactive positions are zero-filled:
\begin{equation}
    \overline{\mathbf h}_{i,t,e}
    =
    \begin{cases}
        \mathbf h_{i,t,e}, & m'_{i,t,e}=1,\\
        \mathbf 0, & m'_{i,t,e}=0.
    \end{cases}
    \label{eq:zero_fill}
\end{equation}
Given expert order $(e_1,\ldots,e_{|\mathcal E|})$, cooperative fusion is
\begin{align}
    \boldsymbol\beta_{i,t}
    &=\operatorname{MLP}_g
    (\overline{\mathbf h}_{i,t,e_1}\Vert\cdots\Vert
    \overline{\mathbf h}_{i,t,e_{|\mathcal E|}}),
    \nonumber\\
    g_{i,t,e}
    &=m'_{i,t,e}\sigma(\beta_{i,t,e}),
    \nonumber\\
    \mathbf z_{i,t}
    &=\sum_{e\in\mathcal E}
    g_{i,t,e}\overline{\mathbf h}_{i,t,e}.
    \label{eq:sparse_fusion}
\end{align}
Unlike competitive softmax routing, sigmoid fusion allows coexisting components such as IP, TCP, and TLS to contribute simultaneously.

\subsection{Flow-Level Pretraining and Classification}
\label{subsec:flow_objectives}

Packets from the same flow describe one communication process.
PTT-Full therefore defines contrastive relations at the flow level rather than treating packets as independent samples.
During pretraining, two views perturb node observations and selected auxiliary sink edges while sharing the canonical PTGs, expert assignments, and presence masks.
These temporary perturbations do not change the schemas later used for compression.

The sparse backbone produces packet representations $\mathbf z_{i,t}^{(1)}$ and $\mathbf z_{i,t}^{(2)}$.
They are pooled within each flow:
\begin{align}
    \mathbf r_i^{(\xi)}
    &=
    \mathop{\operatorname{MaxPool}}\limits_{(i,t)\in\mathcal I_i^{\mathrm{pkt}}}
    \mathbf z_{i,t}^{(\xi)},
    \qquad \xi\in\{1,2\},
    \nonumber\\
    \mathcal L_{\mathrm{SSL}}
    &=\operatorname{NTXent}
    \left(
        \{\mathbf r_i^{(1)},\mathbf r_i^{(2)}
        \mid i\in\mathcal I_{\mathcal B}^{\mathrm{flow}}\}
    \right).
    \label{eq:compact_ssl}
\end{align} 
Here, $\operatorname{NTXent}$ denotes the normalized
temperature-scaled cross-entropy (NT-Xent) loss. 
The two views of the same flow form a positive pair, while representations from other flows in the macro-batch form negatives. 
Packets from the same flow are never treated as negatives. 

For supervised adaptation, the pretrained backbone is jointly fine-tuned with a classifier: 
\begin{equation}
    \boldsymbol\ell_{i,t}
    =\operatorname{MLP}_{c}(\mathbf z_{i,t}),
    \qquad \boldsymbol\ell_{i,t}\in\mathbb R^C.
\end{equation}
Packet logits are averaged within each flow:
\begin{equation}
    \overline{\boldsymbol\ell}_i
    =
    \frac{1}{T_i}
    \sum_{(i,t)\in\mathcal I_i^{\mathrm{pkt}}}
    \boldsymbol\ell_{i,t}.
    \label{eq:flow_logits}
\end{equation}
The supervised objective is
\begin{equation}
    \mathcal L_{\mathrm{Full}}^{(\mathrm{sup})}
    =
    \mathbb E_{(F_i,y_i)\sim\widehat{\mathcal D}_{\mathrm{tr}}}
    \left[
        \operatorname{CE}(\overline{\boldsymbol\ell}_i,y_i)
    \right],
    \label{eq:flow_supervision}
\end{equation}
where $\operatorname{CE}$ denotes the cross-entropy loss, and $\widehat y_i=\arg\max_c\overline\ell_{i,c}$ at inference. 

Let $\theta_F^*$ be the optimized parameters and
\begin{align}
    \mathcal G_F
    &=\{\mathcal G_{F,e}\mid e\in\mathcal E\},
    \nonumber\\
    M_F
    &=(\theta_F^*,\mathcal G_F,d_F).
    \label{eq:full_teacher}
\end{align}
The trained PTT-Full serves as both the high-capacity classifier and the source of field salience, PTG structure, and parameters for compression. 

\section{PTG-Native Structural Compression}
\label{sec:distillation}

\begin{figure}[t]
    \centering
    \includegraphics[width=\linewidth]{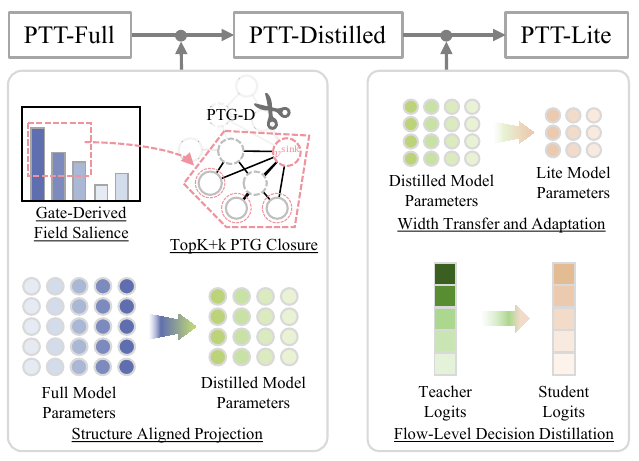}
    \caption{PTG-native structural compression from PTT-Full to PTT-Distilled and PTT-Lite.}
    \label{fig:distillation}
\end{figure}

Sparse dispatch skips absent experts, but each active expert still processes a complete PTG. 
As shown in Fig.~\ref{fig:distillation}, PTT-Distilled compresses these PTGs by selecting salient fields and applying TopK+$k$ closure to preserve their structural context, while PTT-Lite inherits the compressed topology and further reduces representation width. 

\subsection{PTT-Full to PTT-Distilled}
\label{subsec:full_to_distilled}

\paragraph{Gate-Derived Field Salience}
The gates learned by PTT-Full provide a shared salience score for each real protocol field.
For expert $e$, the candidate set is
\begin{equation}
    \mathcal C_e=\mathcal V_{F,e}^{\mathrm{real}}.
    \label{eq:ngi_candidates}
\end{equation}
We normalize the trained gates within each expert to obtain Normalized Field Salience (NFS):
\begin{equation}
    \operatorname{NFS}_e(v)
    =
    \frac{s_{e,v}^{*}}
    {\sum_{u\in\mathcal C_e}s_{e,u}^{*}},
    \qquad v\in\mathcal C_e.
    \label{eq:ngi}
\end{equation}
NFS expresses the learned gate values on a relative scale within each expert and is used only for field ranking. 
Unlike causal attribution methods, NFS is designed as a compression-oriented criterion. 
Its validity is evaluated by whether the resulting protocol-closed subgraphs preserve classification performance under matched compression budgets. 
Given a per-expert budget $K$, we construct the Top-$K$ salience set as: 
\begin{align}
    K_e
    &=\min(K,|\mathcal C_e|),
    \nonumber\\
    \mathcal S_e^K
    &=\operatorname{TopK}_{v\in\mathcal C_e}
    (\operatorname{NFS}_e(v),K_e).
    \label{eq:ngi_topk}
\end{align}

\paragraph{TopK+$k$ PTG Closure}

Retaining the Top-K fields answers what evidence should be preserved, but not how these fields remain meaningful after compression. 
We therefore design TopK+$k$ closure to restore the minimal protocol hierarchy required by these fields. 

Although the PTG is undirected, its hierarchical edges retain the parent--child provenance inherited from packet dissection. 
We represent this provenance by $\operatorname{Par}_{F,e}(v)$, the parent set of node $v$ in the original protocol hierarchy. 
As shown in Algorithm~\ref{alg:topk_closure}, starting from $\mathcal S_e^K$, PTT recursively follows these parent relations until all required ancestors are retained.
Let $\mathcal V_e^{\mathrm{core}}$ denote the selected fields together with the PTG schema nodes recovered by this closure.

The additional closure size is 
\begin{equation}
    k_e
    =
    |\mathcal V_e^{\mathrm{core}}
    \setminus
    \mathcal S_e^K|,
    \qquad
    |\mathcal V_e^{\mathrm{core}}|
    =
    K_e+k_e.
    \label{eq:closure_k}
\end{equation}
The auxiliary sink is added only to preserve the global aggregation structure of the PTG and does not contribute to $k_e$, giving $|\mathcal V_{D,e}|=K_e+k_e+1$. 

\begin{algorithm}[t]
\caption{TopK+$k$ PTG Closure}
\label{alg:topk_closure}
\KwIn{
Full PTGs $\{\mathcal G_{F,e}\}_{e\in\mathcal E}$;
parent maps $\{\operatorname{Par}_{F,e}\}_{e\in\mathcal E}$;
selected field sets $\{\mathcal S_e^K\}_{e\in\mathcal E}$
}
\KwOut{
Distilled PTGs $\mathcal G_D$;
closure sizes $\{k_e\}_{e\in\mathcal E}$
}

$\mathcal G_D \leftarrow \varnothing$\;

\ForEach{$e\in\mathcal E$}{
    $\mathcal V_e^{\mathrm{core}}
    \leftarrow
    \mathcal S_e^K$\;
    
    $\mathcal Q_e
    \leftarrow
    \mathcal S_e^K$\;

    \While{$\mathcal Q_e\neq\varnothing$}{
        select and remove one node $v$ from $\mathcal Q_e$\;

        \ForEach{$u\in\operatorname{Par}_{F,e}(v)$}{
            \If{$u\notin\mathcal V_e^{\mathrm{core}}$}{
                $\mathcal V_e^{\mathrm{core}}
                \leftarrow
                \mathcal V_e^{\mathrm{core}}\cup\{u\}$\;

                $\mathcal Q_e
                \leftarrow
                \mathcal Q_e\cup\{u\}$\;
            }
        }
    }

    $k_e
    \leftarrow
    |\mathcal V_e^{\mathrm{core}}
    \setminus\mathcal S_e^K|$\;

    $\mathcal V_{D,e}
    \leftarrow
    \mathcal V_e^{\mathrm{core}}
    \cup
    \{v_{F,e}^{\mathrm{sink}}\}$\;

    $\mathcal{E}_{D,e}^{\mathrm{hier}}
    \leftarrow
    \{(u,v)\in\mathcal{E}_{F,e}^{\mathrm{hier}}
    \mid
    u,v\in\mathcal V_e^{\mathrm{core}}\}$\;

    $\mathcal{E}_{D,e}^{\mathrm{sink}}
    \leftarrow
    \{(v,v_{F,e}^{\mathrm{sink}})
    \mid
    v\in\mathcal V_e^{\mathrm{core}}\}$\; 

    $\mathcal{E}_{D,e}
    \leftarrow
    \mathcal{E}_{D,e}^{\mathrm{hier}}
    \cup
    \mathcal{E}_{D,e}^{\mathrm{sink}}$\;

    $\mathcal G_{D,e}
    \leftarrow
    (\mathcal V_{D,e},\mathcal{E}_{D,e})$\;

    $\mathcal G_D
    \leftarrow
    \mathcal G_D\cup\{\mathcal G_{D,e}\}$\;
}

\Return{$\mathcal G_D,\{k_e\}_{e\in\mathcal E}$}\;
\end{algorithm}

Ignoring the auxiliary sink node, $\mathcal V_e^{\mathrm{core}}$ is the unique minimal hierarchy-preserving node set that contains $\mathcal S_e^K$ and is closed under the parent relation inherited from the original protocol hierarchy. 
Any smaller node set would necessarily remove either a selected field or an ancestor required by one of its original protocol paths. 
The value of $k_e$ is therefore determined by the protocol paths of the selected fields rather than by a fixed pruning ratio.

Since $\mathcal V_e^{\mathrm{core}}$ prevents a node from being enqueued more than once, the ancestor traversal visits each discovered node once.
Constructing $\mathcal E_{D,e}^{\mathrm{hier}}$ requires one scan of the Full hierarchical edge set, giving an overall complexity of $\mathcal O(|\mathcal V_{F,e}|+|\mathcal E_{F,e}^{\mathrm{hier}}|)$ for each expert. 

\paragraph{Structure-Aligned Projection} 
The PTG-D topology is fixed before parameter projection.
For $d_D<d_F$, the Distilled model is initialized as
\begin{align}
    \theta_D^{(0)}
    &=\Pi_{F\rightarrow D}(\theta_F^*;\mathcal G_D),
    \nonumber\\
    M_D^{(0)}
    &=(\theta_D^{(0)},\mathcal G_D,d_D).
    \label{eq:full_distilled_projection}
\end{align}
Here, $\Pi_{F\rightarrow D}$ denotes structure-aligned channel selection. 
Specifically, $\Pi_{F\rightarrow D}$ selects the top-$d_D$ channels by aggregated squared-$\ell_2$ weight magnitude and slices them consistently across aligned model modules, while retained PTG nodes are matched by protocol-field name. 
The model is then adapted with flow-level supervision:
\begin{align}
    \mathcal L_{\mathrm{Distilled}}^{(\mathrm{sup})}
    &=
    \mathbb E_{(F_i,y_i)\sim\widehat{\mathcal D}_{\mathrm{tr}}}
    [\operatorname{CE}(\overline{\boldsymbol\ell}_{D,i},y_i)],
    \nonumber\\
    \theta_D^*
    &=\arg\min_{\theta_D}
    \mathcal L_{\mathrm{Distilled}}^{(\mathrm{sup})},
    \nonumber\\
    M_D
    &=(\theta_D^*,\mathcal G_D,d_D).
    \label{eq:distilled_training}
\end{align}
PTT-Distilled retains the expert set, anchor experts, and sparse dispatch rule of PTT-Full, but each active expert processes a smaller graph.

\subsection{PTT-Distilled to PTT-Lite}
\label{subsec:distilled_to_lite}

Unlike the previous stage, this stage removes no protocol field. 
PTT-Lite keeps $\mathcal G_D$ unchanged and only reduces representation width from $d_D$ to $d_L<d_D$. 

\paragraph{Width Transfer and Adaptation}
Because PTT-Distilled and PTT-Lite share the same PTG-D schemas and expert organization, their node-specific parameters are structurally aligned: 
\begin{align}
    \theta_L^{(0)}
    &=\Pi_{D\rightarrow L}(\theta_D^*;\mathcal G_D),
    \nonumber\\
    M_L^{(0)}
    &=(\theta_L^{(0)},\mathcal G_D,d_L).
    \label{eq:distilled_lite_projection}
\end{align}
$\Pi_{D\rightarrow L}$ applies the same criterion to select the top-$d_L$ channels and slices the corresponding dimensions consistently across all aligned modules. 
The projected experts are briefly frozen while the narrower fusion and classifier adapt to the new feature space. 
All parameters are then jointly fine-tuned using 
\begin{equation}
    \mathcal L_{\mathrm{Lite}}^{(\mathrm{sup})}
    =
    \mathbb E_{(F_i,y_i)\sim\widehat{\mathcal D}_{\mathrm{tr}}}
    [\operatorname{CE}(\overline{\boldsymbol\ell}_{L,i},y_i)].
    \label{eq:lite_supervision}
\end{equation}

\paragraph{Flow-Level Logits Distillation}
Because the Lite model has lower capacity, it is additionally guided by the flow-level predictions of PTT-Distilled.
With temperature $\tau_{\mathrm{KD}}$,
\begin{equation}
    q_X(c\mid F_i)
    =
    \frac{\exp(\overline\ell_{X,i}^{c}/\tau_{\mathrm{KD}})}
    {\sum_{c'=1}^{C}\exp(\overline\ell_{X,i}^{c'}/\tau_{\mathrm{KD}})},
    \qquad X\in\{D,L\}.
    \label{eq:flow_kd_distributions}
\end{equation}
The Distilled teacher is fixed during Lite adaptation, and
\begin{equation}
    \mathcal L_{\mathrm{KD}}
    =
    \tau_{\mathrm{KD}}^2
    \mathbb E_{F_i\sim\widehat{\mathcal D}_{\mathrm{tr}}}
    \left[
        \operatorname{KL}
        (q_D(\cdot\mid F_i)\Vert q_L(\cdot\mid F_i))
    \right].
    \label{eq:flow_kd_loss}
\end{equation}
The final objective is
\begin{align}
    \mathcal L_{\mathrm{Lite}}
    &=(1-\lambda_{\mathrm{KD}})
    \mathcal L_{\mathrm{Lite}}^{(\mathrm{sup})}
    +\lambda_{\mathrm{KD}}\mathcal L_{\mathrm{KD}},
    \nonumber\\
    \theta_L^*
    &=\arg\min_{\theta_L}\mathcal L_{\mathrm{Lite}},
    \nonumber\\
    M_L
    &=(\theta_L^*,\mathcal G_D,d_L).
    \label{eq:lite_objective}
\end{align}
This loss transfers flow-level decisions without changing the PTG-D topology.

The inherited PTG backbone represents each protocol field with a persistent schema node whose identity is shared across packets.
PTT further uses this field-level alignment as a compression interface: learned node gates determine which protocol evidence is retained, while the protocol hierarchy determines how the retained evidence forms a valid reduced graph. 
This differs from compressing anonymous hidden channels or independently removing graph nodes, because every structural decision remains associated with a concrete protocol field and its original context.
The same field identities are preserved across PTT-Full, PTT-Distilled, and PTT-Lite, which also provides a natural alignment for cross-stage parameter transfer. 

\section{Evaluation}
\label{sec:evaluation}

\subsection{Experimental Setup}
\label{subsec:exp_setup}

\paragraph{Datasets and Evaluation Protocol}
We evaluate PTT on CSTNET-TLS1.3, which contains encrypted sessions across 26 domains, and CipherSpectrum, which contains 120,000 TLS~1.3 sessions across 41 domains and three cipher suites.
All methods use identical flow-disjoint splits and SII-masked packet traces. 
The complete ETH layer, IP addresses, transport ports, and server names are removed before constructing each model's input representation. 
PTT uses no composite TLS fingerprints such as JA3/JA4. 
Retained TLS fields remain individual nodes in their protocol hierarchy. 

\paragraph{Compared Methods}
We compare PTT-Full, PTT-Distilled, and PTT-Lite with image-like, byte/sequence, and graph classifiers, including RF\cite{Method_Imagelike1_RF}, YaTC\cite{Method_Imagelike2_YaTC}, TMWF\cite{Method_Sequence1_TMWF}, BAPM\cite{Method_Sequence2_BAPM}, Var-CNN\cite{Method_Sequence3_VarCNN}, DF\cite{Method_Sequence4_DF}, Tik-Tok\cite{Method_Sequence5_TikTok}, AWF\cite{Method_Sequence6_AWF}, Deep-Packet\cite{Method_sequence9_deeppacket}, ET-BERT\cite{Method_Sequence7_ETBERT}, RBLJAN\cite{related_byte1_RBLJAN}, GNN, and GAT. 
PTGAMoE\cite{PTGAMoE}, which uses dense PTG experts and supervised training, serves as the direct architectural baseline. 
All baselines are retrained under the same flow-disjoint splits and SII-masked inputs, rather than using their originally reported results. 

\paragraph{Metrics and Inference Measurement}
Accuracy and Macro-F1 measure classification performance. 
We additionally report parameter count (Params), effective GFLOPs (GFLOPs), CPU p50 latency (Lat.), and flow throughput (flow/s) to evaluate efficiency. 
Unless stated otherwise, GFLOPs count only the experts actually executed.
Latency is measured over one macro-batch of $N_{\mathcal B}=32$ flows, each retaining at most $P=32$ packets. 
Packet parsing, protocol dissection, and input construction are excluded for all methods. 
For PTT, the timed forward pass includes gathering, sparse dispatch, graph execution, scattering, expert fusion, and classification. 
Flow throughput is computed as $\mathrm{Flows/s}=\frac{N_{\mathcal B}}{\mathrm{Latency}/1000}$, where latency is measured in milliseconds per macro-batch. 
All models are benchmarked on the same CPU with the same macro-batch size, packet limit, and timing boundary. 

\paragraph{Implementation Details}
Code is available at \url{https://anonymous.4open.science/r/pruned_traffic_trees-D08C}. 
Training uses a server with Intel i5-13490F CPU, 32\,GB memory, and an NVIDIA RTX 4060 Ti 16\,GB GPU. 
CPU inference uses a server with Intel i5-9500T and 8\,GB memory.
PTT-Full uses IP, TCP-core, TCP-option, TLS-record-header, and TLS-record-payload experts with $d_F=128$. 
Unless otherwise stated, PTT-Distilled uses $K=5$ and $d_D=32$, and PTT-Lite uses $d_L=20$. 
The Lite KD weights are 0.10 on CSTNET-TLS1.3 and 0.05 on CipherSpectrum. 

\subsection{Overall Performance and CPU Efficiency}
\label{subsec:overall_performance_and_cpu_efficiency}

\begin{table*}[t]
\centering
\caption{Overall comparison on CSTNET-TLS1.3 and CipherSpectrum.}
\label{tab:classification_performance_overall}
\setlength{\tabcolsep}{5.0pt}
\renewcommand{\arraystretch}{1.08}
\begin{tabular}{ll@{\quad}|@{\quad}cccc@{\quad}|@{\quad}cccc}
\toprule[1.2pt]
\multicolumn{2}{c@{\quad}|@{\quad}}{Dataset}
& \multicolumn{4}{c@{\quad}|@{\quad}}{CSTNET-TLS1.3}
& \multicolumn{4}{c}{CipherSpectrum} \\
\midrule
Category & Method
& AC & F1 & Params (M) & Lat. (ms/batch)
& AC & F1 & Params (M) & Lat. (ms/batch) \\
\midrule
\multirow{2}{*}{\textit{Image-like}}
& RF
& 0.3648 & 0.2996 & 0.924 & 4.73
& 0.6251 & 0.6318 & 0.947 & 4.41 \\
& YaTC
& 0.8756 & 0.8599 & 1.863 & 39.72
& 0.1162 & 0.1146 & 1.866 & 41.49 \\
\midrule
\multirow{9}{*}{\textit{Byte/Sequence}}
& TMWF
& 0.2953 & 0.2026 & 4.834 & 28.62
& 0.5289 & 0.5175 & 4.838 & 28.24 \\
& BAPM
& 0.4069 & 0.3505 & 0.238 & 1.48
& 0.5097 & 0.5028 & 0.269 & 1.49 \\
& Var-CNN
& 0.4963 & 0.4925 & 8.766 & 23.89
& 0.2455 & 0.2072 & 8.782 & 23.27 \\
& DF
& 0.3449 & 0.2555 & 3.679 & 5.25
& 0.5949 & 0.5913 & 3.687 & 5.20 \\
& Tik-Tok
& 0.2233 & 0.1378 & 3.679 & 5.20
& 0.7176 & 0.7135 & 3.687 & 5.25 \\
& AWF
& 0.2655 & 0.1908 & 0.048 & 0.59
& 0.3640 & 0.3389 & 0.069 & 0.59 \\
& Deep-Packet
& 0.0656 & 0.0305 & 10.108 & 2.67
& 0.0563 & 0.0026 & 10.109 & 2.40 \\
& ET-BERT
& 0.7761 & 0.7759 & 132.146 & 304.81
& 0.4059 & 0.3321 & 132.157 & 303.88 \\
& RBLJAN
& 0.8266 & 0.7859 & 0.357 & 60.44
& 0.5808 & 0.5427 & 0.499 & 74.88 \\
\midrule
\multirow{3}{*}{\textit{Graph}}
& GNN
& 0.8530 & 0.8230 & 0.026 & 1.51
& 0.7888 & 0.7880 & 0.029 & 1.37 \\
& GAT
& 0.8905 & 0.8662 & 0.141 & 1.61
& 0.8337 & 0.8315 & 0.144 & 1.46 \\
& PTGAMoE
& 0.9395 & 0.9265 & 0.962 & 1001.93
& 0.8720 & 0.8709 & 1.263 & 1150.57 \\
\midrule
\multirow{3}{*}{\textit{PTT}}
& PTT-Full
& \textbf{0.9646} & \textbf{0.9519} & 0.943 & 316.20
& \textbf{0.9415} & \textbf{0.9416} & 1.243 & 328.66 \\
& PTT-Distilled
& \underline{0.9472} & \underline{0.9380} & 0.218 & 41.79
& \underline{0.9241} & \underline{0.9238} & 0.513 & 48.31 \\
& PTT-Lite
& 0.9424 & 0.9325 & 0.186 & 36.12
& 0.9137 & 0.9136 & 0.481 & 38.85 \\
\bottomrule[1.2pt]
\end{tabular}

\end{table*}

Table~\ref{tab:classification_performance_overall} shows that PTT-Full achieves the highest Macro-F1 on both datasets.
PTT-Lite retains 0.9325 and 0.9136 Macro-F1 with 0.186M and 0.481M parameters, exceeding GAT by 0.0663 and 0.0821, respectively.
Compared with dense PTGAMoE, PTT-Full improves Macro-F1 by 0.0254 and 0.0707 while reducing macro-batch latency by 3.17$\times$ and 3.50$\times$. 

\begin{figure}[t]
    \centering
    \begin{subfigure}[t]{0.24\textwidth}
        \centering
        \includegraphics[width=\linewidth]{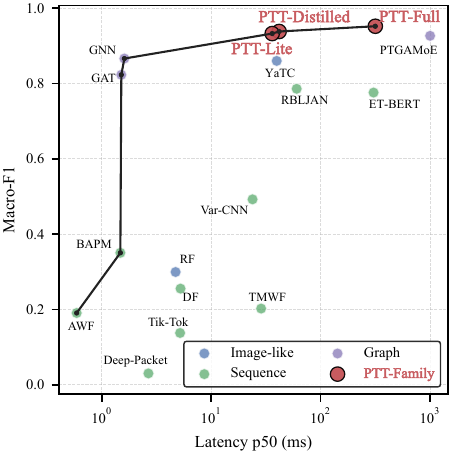}
        \caption{CSTNET-TLS1.3.}
        \label{fig:pareto_cstnet_latency}
    \end{subfigure}
    \hfill
    \begin{subfigure}[t]{0.24\textwidth}
        \centering
        \includegraphics[width=\linewidth]{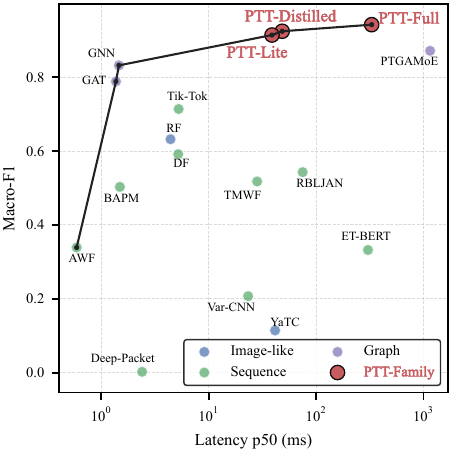}
        \caption{CipherSpectrum.}
        \label{fig:pareto_cipher_latency}
    \end{subfigure}
    \caption{Pareto frontiers of performance--latency trade-offs.}
    \label{fig:pareto_frontier}
\end{figure}

To jointly compare performance and latency, Fig.~\ref{fig:pareto_frontier} plots the performance--latency trade-off leveraging Pareto frontiers. 
A model is non-dominated when no alternative has both higher Macro-F1 and lower latency. 
The PTT variants occupy the high-performance region on both datasets and provide distinct Full, Distilled, and Lite operating points. 

\subsection{Compression Effectiveness}
\label{subsec:compression_effectiveness}

\begin{figure}[t]
    \centering
    \includegraphics[width=\columnwidth]{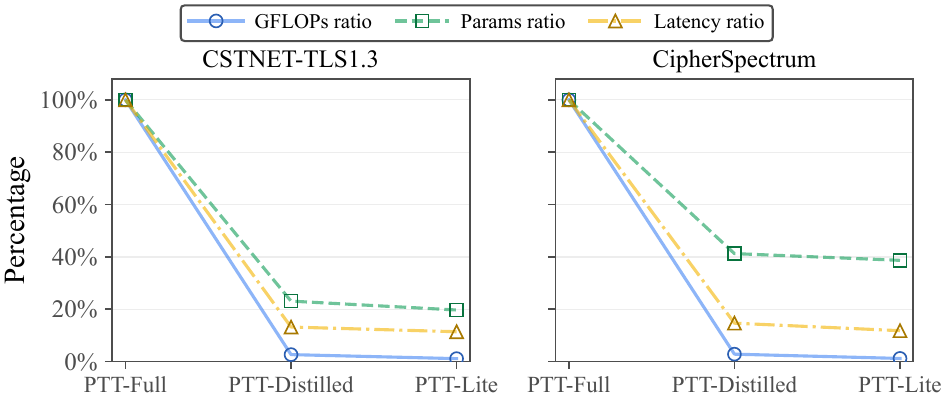}
    \caption{Compression effectiveness of PTT Family.}
    \label{fig:compression}
\end{figure}

As shown in Fig.~\ref{fig:compression}, the Full-to-Distilled transition, which jointly reduces PTG topology and representation width, accounts for most of the overall resource reduction. 
Table~\ref{tab:generic_compression_baselines} further isolates the advantage of PTG-native structural compression under matched computation budgets. 
PTT-Lite keeps the distilled topology unchanged and further narrows the representation width. 
It uses only 19.72\% and 38.70\% of the Full-model parameters and 1.15\% and 1.22\% of its effective GFLOPs. 
The corresponding CPU speedups are 8.75$\times$ and 8.46$\times$, indicating that structural compression provides the main reduction and width compression adds a lighter deployment point. 
This confirms that reducing protocol structures rather than only hidden dimensions is the main source of compression efficiency. 

\subsection{Effect of Presence-Aware Sparse Modeling}
\label{subsec:presence_ssl_ablation}

\begin{table*}[t]
\centering
\caption{Effect of SSL, protocol-presence masking, and actual expert skipping.}
\label{tab:presence_ssl_ablation}
\setlength{\tabcolsep}{4.2pt}
\renewcommand{\arraystretch}{1.10}
\begin{tabular}{l@{\quad}|@{\quad}lccc@{\quad}|@{\quad}lclr}
\toprule
Dataset & ID & SSL & Mask & Sparse & F1 & GFLOPs & Lat. (ms/batch) & Flows/s \\
\midrule
\multirow{5}{*}{\makecell[l]{CSTNET-TLS1.3}}
& D0 & \xmark & \xmark & \xmark & 0.9296 $(\pm 0.0146)$ & 9.241 & 436.04 [28.95] & 73.54 [4.51] \\
& D1 & \cmark & \xmark & \xmark & 0.9301 $(\pm 0.0046)$ & 9.241 & 436.94 [32.34] & 73.57 [4.63] \\
& D2 & \xmark & \cmark & \xmark & 0.9415 $(\pm 0.0042)$ & 9.241 & 436.87 [28.63] & 73.51 [4.52] \\
& D3 & \cmark & \cmark & \xmark & 0.9519 $(\pm 0.0111)$ & 9.241 & 438.35 [31.08] & 73.40 [4.80] \\
& \textbf{D4} & \cmark & \cmark & \cmark & \textbf{0.9519 $(\pm 0.0111)$} & \textbf{6.961} & \textbf{316.20} [18.53] & \textbf{100.28} [6.00] \\
\midrule
\multirow{5}{*}{\makecell[l]{CipherSpectrum}}
& D0 & \xmark & \xmark & \xmark & 0.9322 $(\pm 0.0042)$ & 12.444 & 635.04 [24.82] & 50.17 [2.13] \\
& D1 & \cmark & \xmark & \xmark & 0.9360 $(\pm 0.0074)$ & 12.444 & 641.06 [26.53] & 49.52 [2.04] \\
& D2 & \xmark & \cmark & \xmark & 0.9402 $(\pm 0.0036)$ & 12.444 & 639.01 [25.61] & 49.74 [2.12] \\
& D3 & \cmark & \cmark & \xmark & 0.9416 $(\pm 0.0013)$ & 12.444 & 637.52 [25.44] & 49.96 [2.26] \\
& \textbf{D4} & \cmark & \cmark & \cmark & \textbf{0.9416 $(\pm 0.0013)$} & \textbf{6.371} & \textbf{328.66} [14.82] & \textbf{95.73} [4.51] \\
\bottomrule
\end{tabular}

\vspace{0.5ex}
\footnotesize
\emph{Note:} Latency and throughput are median [IQR] over 15 runs: three seeds and five repetitions per seed. 
\end{table*}

Table~\ref{tab:presence_ssl_ablation} separates the modeling effect of protocol-presence masking from the efficiency gain of actual sparse execution. 
Adding SSL alone (D1) provides limited gains over D0, whereas presence masking (D2) improves Macro-F1 and further combines with SSL (D3) to reach 0.9519 and 0.9416. 
D3 and D4 share the same checkpoints and predictions. 
D4 only enables sparse execution for absent protocol components. 
It reduces GFLOPs by 24.7\% and 48.8\%, latency by 27.9\% and 48.4\%, and increases throughput by 36.6\% and 91.6\%. 
Thus, presence masking improves protocol-aware modeling, while sparse execution turns protocol absence into computation savings without changing predictions. 

\subsection{Effect of PTG-Native Structural Compression}
\label{subsec:ablation_structure}

This section examines four questions: 
(1) whether PTG-native compression is more effective than model-level compression; 
(2) whether learned field salience guides selection; 
(3) whether retained fields require their native hierarchy; 
and (4) how the field budget controls PTG-D size. 

\paragraph{Model-Level versus PTG-Native Compression}

We first compare PTT with three validation-tuned model-level baselines that retain the complete PTG. 
Narrow-CE (N-CE) reduces model width and trains the resulting full-PTG model with cross-entropy, while Narrow-KD (N-KD) additionally transfers PTT-Full logits. 
As a structured pruning baseline, $\ell_2$-based channel pruning (L2-Ch) ranks channels by the aggregated $\ell_2$ norm of their associated weights and retains the highest-ranked channels before validation-tuned adaptation. 

To ensure compute-matched comparison, we enumerate full-PTG widths $d\in\{4,8,12,16,20,24,28,32\}$ and select the largest width whose validation-set effective GFLOPs do not exceed the corresponding PTT-Distilled or PTT-Lite budget.
This yields $d=20$ and $d=12$, respectively, on both datasets. 
All baselines further tune the learning rate on $\{10^{-4},2\times10^{-4},5\times10^{-4}\}$, while N-KD and L2-Ch also tune their KD weights using validation data. 
The checkpoint with the highest validation Macro-F1 is then evaluated on the test set. 

\begin{table}[t]
\centering
\caption{Validation-tuned model-level compression versus PTT under
upper-bounded Distilled and Lite compute budgets.}
\label{tab:generic_compression_baselines}
\setlength{\tabcolsep}{3.6pt}
\renewcommand{\arraystretch}{1.06}
\resizebox{\columnwidth}{!}{
\begin{tabular}{
ll@{\quad}|@{\quad}
ccc@{\quad}|@{\quad}
ccc
}
\toprule[1.2pt]
\multicolumn{2}{c@{\quad}|@{\quad}}{}
& \multicolumn{3}{c@{\quad}|@{\quad}}{Distilled Budget}
& \multicolumn{3}{c}{Lite Budget} \\
\midrule
Dataset & Method
& F1 & GFLOPs & Lat.
& F1 & GFLOPs & Lat. \\
\midrule

\multirow{4}{*}{\makecell[l]{CSTNET\\TLS1.3}}
& N-CE
& 0.9160 & 0.177 & 76.34
& 0.8261 & 0.075 & 65.38 \\

& N-KD
& 0.8867 & 0.177 & 75.86
& 0.8484 & 0.075 & 65.40 \\

& L2-Ch
& 0.9040 & 0.177 & 75.88
& 0.8921 & 0.075 & 64.85 \\

& \textbf{PTT}
& \textbf{0.9380} & 0.187 & \textbf{41.79}
& \textbf{0.9325} & 0.080 & \textbf{36.12} \\

\midrule

\multirow{4}{*}{\makecell[l]{Cipher\\Spectrum}}
& N-CE
& 0.8914 & 0.173 & 88.25
& 0.7712 & 0.073 & 70.32 \\

& N-KD
& 0.8589 & 0.173 & 87.69
& 0.7749 & 0.073 & 70.34 \\

& L2-Ch
& 0.8850 & 0.173 & 87.72
& 0.8393 & 0.073 & 69.74 \\

& \textbf{PTT}
& \textbf{0.9238} & 0.182 & \textbf{48.31}
& \textbf{0.9136} & 0.078 & \textbf{38.85} \\

\bottomrule[1.2pt]
\end{tabular}
}
\end{table}

Table~\ref{tab:generic_compression_baselines} shows that PTT consistently retains more classification performance than the strongest validation-tuned model-level baseline. 
Its Macro-F1 gains are 0.0220 and 0.0324 at the Distilled budget, and increase to 0.0404 and 0.0743 at the Lite budget on CSTNET-TLS1.3 and CipherSpectrum, respectively. 
Although the full-PTG baselines use slightly fewer effective GFLOPs, their CPU latency is approximately $1.83\times$ higher at the Distilled budget and $1.80\times$ higher at the Lite budget. 
These results indicate that removing less informative protocol structure provides a better performance--efficiency trade-off than retaining the complete PTG and compressing model width or channels alone. 

\paragraph{Field Guidance and Salience Stability}
We next examine which real protocol fields should be retained before
closure. 
Random (Rand.) samples candidate fields uniformly, Frequency (Freq.)
ranks them by observation frequency in the training set, Taylor~\cite{TaylorPruning} uses
first-order training-loss sensitivity, and NFS ranks fields according to
the gates learned by PTT-Full. 
Because the NFS normalization is shared by all candidates within the
same expert, it does not change their ordering. 
The resulting Top-$K$ selection is determined by the learned gate salience. 
All methods use the same $K=5$ and TopK+$k$ closure. 

\begin{figure}[t]
    \centering
    \includegraphics[width=\columnwidth]{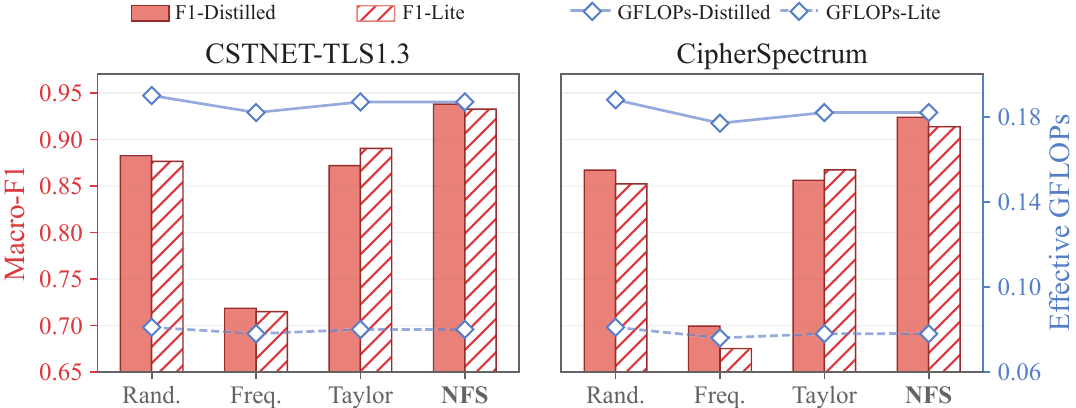}
    \caption{Field-selector comparison under the same TopK+$k$ closure and $K=5$.}
    \label{fig:selector}
\end{figure}

As shown in Fig.~\ref{fig:selector}, the learned gate salience produces the strongest compressed models under comparable graph-computation budgets.
We further examine the stability of the selected fields across three independent training runs.
For experts with more than five candidate fields, the selected Top-$5$ sets achieve average pairwise Jaccard similarities of $0.4286\pm0.1431$ on CSTNET-TLS1.3 and $0.3982\pm0.1536$ on CipherSpectrum.
The similar overlaps on both datasets indicate that the exact Top-$K$ selections remain seed-sensitive rather than invariant. 
Nevertheless, repeatedly selected protocol fields form a more stable semantic core, while fields near the selection boundary may vary. 
Specifically, on CSTNET-TLS1.3, several fields, including IP length, TCP length, TLS cipher suite, and TLS extension type, are consistently retained across all seeds. 
The learned salience therefore provides a useful compression signal with a stable semantic core rather than a deterministic ranking of every protocol field. 

\paragraph{Protocol Closure and Case Study}
After fixing the selected fields, we examine whether their original hierarchical relations should also be retained. 
Selected Fields Only (SFO) keeps only the selected real fields. 
Selected + Ancestors (S+A) additionally retains the same ancestor nodes used by TopK+$k$, but removes their original parent--child relations. 
TopK+$k$ retains both the required ancestors and their original hierarchical edges. 
Path completeness equals one only when every node and hierarchical edge on the original Full-PTG path of a selected field is retained. 

\begin{table}[t]
\centering
\caption{PTG-D retention schemes using the same NFS-selected fields
and $K=5$. Here, `A.N.' denotes ancestor nodes, `H.E.' denotes hierarchical edges, and `Comp.' denotes path completeness, respectively. }
\label{tab:closure_comparison}
\setlength{\tabcolsep}{5.2pt}
\renewcommand{\arraystretch}{1.08}
\begin{tabular}{ll@{\quad}|@{\quad}ccc@{\quad}|@{\quad}cc}
\toprule[1.2pt]
Dataset & Scheme
& A.N. & H.E. & Comp.
& F1-D & F1-L \\
\midrule

\multirow{3}{*}{\makecell[l]{CSTNET\\TLS1.3}}
& SFO
& \xmark
& \xmark
& 0.00
& 0.9010
& 0.8894 \\

& S+A
& \cmark
& \xmark
& 0.00
& 0.8834
& 0.8636 \\

& \textbf{TopK+$k$}
& \cmark
& \cmark
& \textbf{1.00}
& \textbf{0.9380}
& \textbf{0.9325} \\

\midrule

\multirow{3}{*}{\makecell[l]{Cipher\\Spectrum}}
& SFO
& \xmark
& \xmark
& 0.00
& 0.8860
& 0.8663 \\

& S+A
& \cmark
& \xmark
& 0.00
& 0.8679
& 0.8380 \\

& \textbf{TopK+$k$}
& \cmark
& \cmark
& \textbf{1.00}
& \textbf{0.9238}
& \textbf{0.9136} \\

\bottomrule[1.2pt]
\end{tabular}
\end{table}

Table~\ref{tab:closure_comparison} shows that retaining additional ancestor nodes alone is insufficient. 
S+A keeps the same ancestor nodes as TopK+$k$ but performs below even SFO when their original hierarchical relations are removed. 
In contrast, TopK+$k$ achieves complete protocol paths and the highest Macro-F1 on both datasets. 
Together with Fig.~\ref{fig:selector}, this result separates the two roles of PTG-native compression: learned field salience determines \emph{what} protocol evidence is retained, while closure determines \emph{how} that evidence remains connected to its native hierarchy. 

\begin{figure}[t]
    \centering
    \includegraphics[width=\columnwidth]{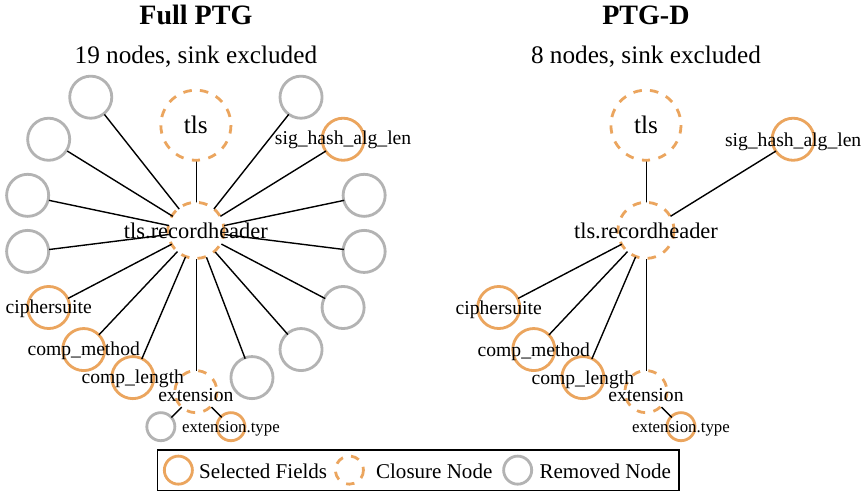}
    \caption{PTG-D construction for the TLS-Record-Header expert. }
    \label{fig:ptgd_case_study}
\end{figure}

Fig.~\ref{fig:ptgd_case_study} gives a concrete example using the TLS-Record-Header expert. 
Five real fields are selected by the learned salience, while closure restores three intermediate protocol nodes required by their original paths. 
For example, \texttt{tls.recordheader.extension.type} is retained together with \texttt{tls}, \texttt{tls.recordheader}, and \texttt{tls.recordheader.extension}, rather than becoming an isolated field. 
The resulting PTG-D reduces this expert from 19 to 8 PTG schema nodes, excluding the fixed sink, while preserving the protocol context of all selected fields. 

\paragraph{Structural Reduction and Field Budget}
We finally measure the actual graph reduction produced by the default $K=5$ setting. 
PTG-D removes more than half of the graph structure on both datasets. 
CSTNET-TLS1.3 is reduced from 82 to 38 nodes and from 72 to 28 hierarchical edges, while CipherSpectrum is reduced from 88 to 38 nodes and from 78 to 28 edges. 
These correspond to node reductions of 53.66\%--56.82\% and edge reductions of 61.11\%--64.10\%, while path completeness remains 1.00. 

Compression is not uniform across experts: node retention ranges from 25.93\% to 88.89\% on CSTNET-TLS1.3 and from 25.93\% to 85.71\% on CipherSpectrum.
PTT therefore fixes a semantic field budget rather than a graph-retention ratio. 
The final PTG-D size is determined by field availability and the protocol paths required by the selected fields. 

\begin{figure}[t]
    \centering
    \includegraphics[width=\columnwidth]{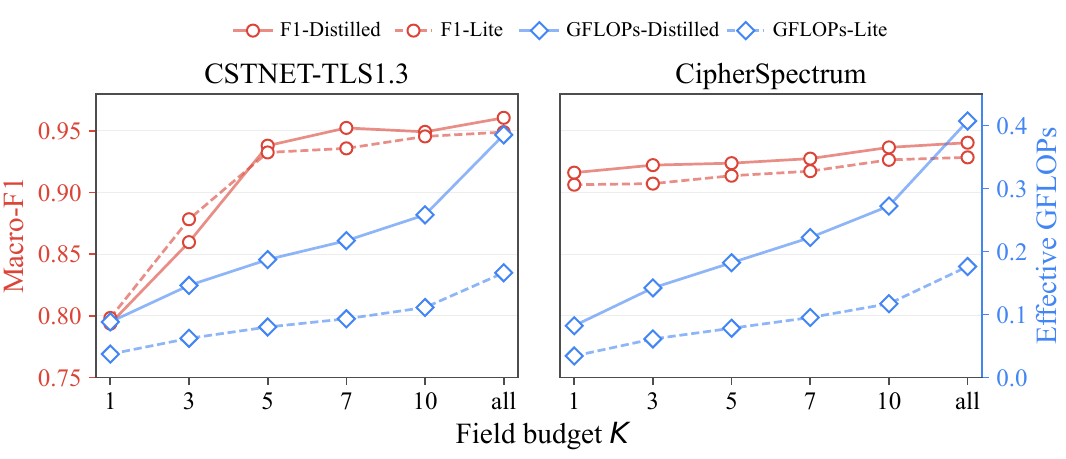}
    \caption{Effect of the field budget $K$ in TopK+$k$ closure.}
    \label{fig:effect_K}
\end{figure}

Fig.~\ref{fig:effect_K} further varies this field budget.
To emphasize the actual performance--computation trade-off, we report Macro-F1 and GFLOPs and omit parameter counts, which change only marginally with $K$. 

On CSTNET-TLS1.3, the performance gain begins to saturate around $K=5$: larger budgets increase graph computation substantially while providing smaller additional gains. 
CipherSpectrum benefits more consistently from additional fields and therefore admits higher-accuracy operating points at $K=10$ or $K=\mathrm{all}$. 
We use $K=5$ as the common compact setting because it substantially reduces graph computation on both datasets while retaining strong classification performance. 

\subsection{Effect of Cross-Stage Transfer}
\label{subsec:ablation_transfer}

After obtaining PTG-D, we further investigate how knowledge is transferred from the structurally compressed teacher to the deployment-oriented Lite model. 
In the following experiments, the target PTG topology and model width are kept unchanged. 

\paragraph{Effect of Channel-Projection Initialization}

\begin{figure}[t]
    \centering
    \begin{subfigure}[t]{0.24\textwidth}
        \centering
        \includegraphics[width=\linewidth]{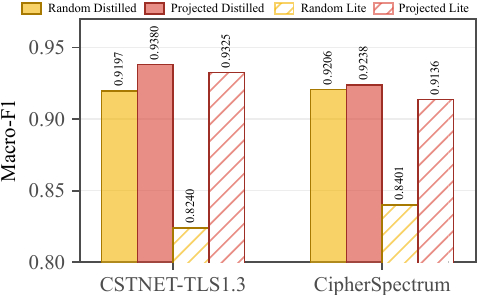}
        \caption{Effect of channel-projection initialization.}
        \label{fig:projection_init}
    \end{subfigure}
    \hfill
    \begin{subfigure}[t]{0.24\textwidth}
        \centering
        \includegraphics[width=\linewidth]{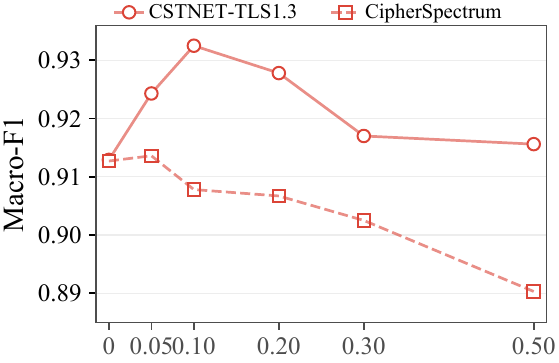}
        \caption{Effect of the logits-KD weight ($\lambda_{\mathrm{KD}}$) for PTT-Lite. }
        \label{fig:effect_kd}
    \end{subfigure}
    \caption{Effect of Cross-Stage Transfer.}
    \label{fig:cross_transfer}
\end{figure}

Fig.~\ref{fig:cross_transfer}(a) compares random initialization with structure-aligned projection. 
Projection improves Full-to-Distilled Macro-F1 by 0.0183 and 0.0032, and yields larger gains of 0.1085 and 0.0735 for Distilled-to-Lite. 
The larger Lite gains indicate that transferred channels mainly stabilize adaptation to an already fixed compact architecture. 

\paragraph{Effect of Flow-Level Logits Distillation}

Fig.~\ref{fig:cross_transfer}(b) evaluates whether flow-level teacher predictions provide additional guidance after width reduction. 
On CSTNET-TLS1.3, $\lambda_{\mathrm{KD}}=0.10$ raises Macro-F1 from 0.9129 to 0.9325. 
CipherSpectrum benefits only marginally at 0.05 and degrades under larger teacher weights. 
Logits distillation is therefore complementary and dataset-dependent rather than the main source of structural-compression gains. 

\paragraph{Effect of Width Selection}
\label{subsec:ablation_width}

\begin{table}[t]
\centering
\caption{Effect of the PTT-Lite hidden dimension $d_L$.}
\label{tab:width_pareto}
\setlength{\tabcolsep}{8.0pt}
\renewcommand{\arraystretch}{1.08}
\begin{tabular}{l@{\quad}|@{\quad}c@{\quad}|@{\quad}ccc}
\toprule
Dataset & $d_L$ & F1 & Params (M) & GFLOPs \\
\midrule
\multirow{3}{*}{\makecell[l]{CSTNET-TLS1.3}}
& 16 & 0.9275 & 0.178 & 0.054 \\
& 20 & 0.9325 & 0.186 & 0.080 \\
& 28 & 0.9318 & 0.206 & 0.146 \\
\midrule
\multirow{3}{*}{\makecell[l]{CipherSpectrum}}
& 16 & 0.8883 & 0.473 & 0.053 \\
& 20 & 0.9136 & 0.481 & 0.078 \\
& 28 & 0.9346 & 0.501 & 0.143 \\
\bottomrule
\end{tabular}
\end{table}

Table~\ref{tab:width_pareto} varies $d_L$ within the fixed PTG-D topology. 
On CSTNET-TLS1.3, increasing $d_L$ from 16 to 20 improves Macro-F1, while $d_L=28$ adds computation without further gain. 
CipherSpectrum needs more capacity: $d_L=16$ degrades clearly, and $d_L=28$ reaches 0.9346 at 0.143 GFLOPs. 
We use $d_L=20$ as a common balanced point, while $d_L=28$ remains an accuracy-oriented option for CipherSpectrum. 

\section{Conclusion}
\label{sec:conclusion}

We present Pruned Traffic Trees (PTT), a protocol-structured ETC model family that makes native protocol structures directly compressible. 
PTT determines what protocol evidence to retain through learned field salience and how to preserve its structural context through TopK+$k$ closure, constructing compact PTG-Ds without breaking native protocol paths. 
Together with sparse execution and cross-stage adaptation, PTT provides practical Full, Distilled, and Lite operating points. 
Experiments under flow-disjoint and SII-masked settings demonstrate strong Macro-F1 with substantially reduced graph computation and CPU latency. 
Future work will investigate end-to-end parsing overhead, optimize graph execution, and extend PTT to broader protocols. 


\bibliographystyle{IEEEtran}
\bibliography{references}

\end{document}